\documentstyle[twoside,fleqn,espcrc2]{article}

\newcommand{\AmS}{{\protect\the\textfont2
  A\kern-.1667em\lower.5ex\hbox{M}\kern-.125emS}}

\title{Towards a lattice calculation of the coefficients of the QCD
chiral Lagrangian\thanks{This research is supported in part under DOE
grant DE-FG02-91ER40676.}}

\author{A.R. Levi, V.Lubicz and C.Rebbi \\
Department of Physics, Boston University, 590 Commonwealth Avenue,
Boston, MA 02215, USA}

\begin{document}

\begin{abstract}
We discuss a general strategy to compute the coefficients of QCD chiral
Lagrangian by using the lattice regularization of QCD with Wilson
fermions. This procedure requires the introduction of an effective
Lagrangian for lattice QCD as an intermediate step in the
calculation. The continuum QCD chiral Lagrangian can be then obtained by
expanding the lattice effective Lagrangian in increasing powers of the
external momenta. A suitable renormalization procedure is required to
account for the chiral symmetry breaking introduced by the Wilson term
in the lattice action. In anticipation of a numerical simulation, the
lattice effective Lagrangian is computed analytically and investigated
in the strong coupling and large $N$ limit. 
\end{abstract}

\maketitle

\section{Introduction}
A first principle calculation of the coefficients of the QCD chiral
Lagrangian \cite{wein,gl} is a task of fundamental theoretical  
interest. Besides representing an important test of QCD in the
low-energy sector, such a calculation would also improve our knowledge
of the chiral coefficients. At present, these coefficients are mainly
determined from phenomenological constraints and are known, in some
cases, with relative uncertainties larger than $100\%$ \cite{beg}.

Being only a function of the fundamental scale $\Lambda_{QCD}$ and
the heavy quark masses, the coefficients of chiral Lagrangian are in
principle directly calculable from QCD. This requires the use of
lattice QCD, which represents today the only viable computational method
for calculating non-perturbative QCD observables from first
principles. The aim of this talk is to discuss a general strategy to
perform such a calculation. 

A standard approach to compute the coefficients of an effective theory
consists in performing a matching between the effective and the
underlying fundamental theory. Specifically, one computes a sufficient 
numbers of physical amplitudes, both in the effective and the original
theory, and derives the values of the coefficients from a comparison of
the results. For the QCD chiral Lagrangian, the calculation in the
fundamental theory, being non-perturbative, should be performed on
the lattice.

However, such an approach would be significantly affected by the
existence, on the lattice, of an infra-red cut-off. In current numerical
simulations, the minimum allowed value of momentum, $p_{min}=2\pi/La$,
is typically of the order of the rho meson mass. This prevents the
possibility of performing the calculation in the low-energy region,
where the predictions of the effective theory can be reliably
obtained. In order to overcome this difficulty, one should then consider
lattices of very large volumes, which unfortunately entails substantial
losses in efficiency and, therefore, in statistical accuracy.

An alternative approach to the calculation of the coefficients of QCD
chiral Lagrangian on the lattice has been proposed in ref.~\cite{cs}.
The basic observation there is that the separation between effective and
non-effective degrees of freedom, occurring in the continuum theory,
must be mirrored by an equivalent distinction in the theory regularized
on the lattice. Therefore, one can define an effective theory on the
lattice which is equivalent, for any value of the lattice spacing or the
lattice bare coupling constant, to the fundamental lattice QCD
theory. The coefficients of the continuum QCD chiral Lagrangian can be
then computed through the following steps:
i) Define the lattice effective theory by assuming a sufficiently
large set of couplings, with strengths determined by arbitrary
coefficients.
ii) Fix these coefficients through the matching of an overcomplete
set of expectation values, computed both in the full and the effective
lattice theory. The calculation in the full theory requires a numerical
simulation.
iii) Expand the effective lattice Lagrangian in increasing powers of the
external momenta. The result of such an expansion is the continuum QCD
chiral Lagrangian. 

The basic feature of this procedure is the introduction of the
effective lattice Lagrangian as an intermediate step in the
calculation. The main advantage is that this Lagrangian can be organized
in terms of the distance of couplings rather then in powers of momenta,
as in the corresponding continuum effective theory. Therefore, in
performing the matching between the effective and the fundamental
theory, the existence of an infra-red cut-off in the calculation does
not represent a problem anymore. 

In addition, in this procedure both the full and the effective theory
are defined in the same regularization scheme. Therefore, the finite
ultraviolet cut-off effects, affecting the numerical calculation, can
be better kept under control, since these effects are exactly predicted
by the lattice effective theory.

Since the effective Lagrangian contains explicitly the collective fields
responsible for the long distance behavior of the fundamental lattice
theory, the calculations of its coefficients should only involve shorter
scale of length, and should be feasible, therefore, on lattices of
moderate size.

\section{The lattice effective Lagrangian in the strong coupling and
large $N$ limit} 

A useful feature of the strategy discussed above is that, in the limit
of strong coupling and large number of colors $N$, the integration of
the non-effective degrees of freedom in the QCD lattice Lagrangian can
be performed analytically \cite{kluks}. Therefore, in this limit, the
corresponding effective theory can be exactly computed and it provides a
useful indication on the general structure of the lattice effective
Lagrangian. Being only a consequence of chiral invariance, charge
conjugation and discrete lattice symmetries, this structure must be
preserved in the region of intermediate gauge couplings, which is
relevant for simulations of continuum QCD.  

In the strong coupling and large $N$ limit, the effective Lagrangian for
lattice QCD with Wilson fermions, and a set of proper external sources,
has the form \cite{llr}:
\begin{equation}
\label{leff}
\begin{array}{l}
{\cal L}_{eff}(U) = \\ C_{\Delta} {\rm Tr } \left[\left(\Delta_ \mu
U\right) ^{\dagger} \left(\Delta _\mu U\right)\right] + C_m {\rm Tr}
\left[ \chi ^{\dagger} U + U^{\dagger } \chi \right] \\ + C_W r^2 
{\rm Tr}\left[\xi ^{\dagger} U \xi ^{\dagger} U + U^{\dagger} \xi U
^{\dagger} \xi \right] + \ldots 
\end{array}
\end{equation}
where the dots represent an infinite series of additional terms. In
eq.~(\ref{leff}), $U$ is the effective field, $\Delta _\mu U$ represents
a lattice version of the chiral covariant derivative, $\chi$ is the
external source including also the light quark mass matrix and $\xi$ is
an additional source which is coupled, in the lattice fermionic action,
to the Wilson term. This source has no correspondent in the continuum
theory, and it is introduced in the lattice action in order to achieve a
local chiral invariance with respect to combined transformations of the
quark fields and the external sources. 

The first two terms on the right-hand side of eq.~(\ref{leff}) formally
reduce in the continuum limit to the operators appearing in the
QCD chiral Lagrangian at ${\cal O}(p^2)$. In contrast, the last term in
eq.~(\ref{leff}) does not have any correspondent in the continuum
effective theory. This term is an example of an infinite number of
couplings which are originated by the Wilson term. They mainly appear,
in the effective Lagrangian, in order to reproduce the effects of the
explicit chiral symmetry breaking induced  by the Wilson term in the
lattice theory.

The role of these ``Wilson terms'' in the effective Lagrangian can be
studied explicitly. A clear example is given, in this respect, by the
well known equation which relates, in the continuum, the pseudoscalar
mass and decay constant to the quark mass and condensate. By using the
effective theory, in the strong coupling and large $N$ limit, we find
that on the lattice this relation has the form: 
\begin{equation}
\label{gor}
Z_A^2 F_\pi^2 M_\pi^2 = \left( m-\widehat{m}\right) \langle \bar\psi \psi
\rangle  
\end{equation}
where $Z_A=1+r^2$ and $\widehat{m}$ is a function of the bare quark mass
$m$ and the Wilson parameter $r$. This function defines the critical
value $m_c$ of the bare quark mass. In the limit $r=0$, chiral symmetry
prevents an additive renormalization of the quark mass; the function
$\widehat{m}$ identically vanishes and $m_c=0$. In the case $r=1$, one
finds that the pion mass vanishes for $m_c=-2$, corresponding to the
critical value $k_c=1/4$ of the Wilson hopping parameter. 

The constant $Z_A$ in eq.~(\ref{gor}) is also a consequence of the
Wilson terms in the effective Lagrangian. Despite the strong coupling
approximation, this constant can be consistently interpreted as a
renormalization of the point-split axial current $A_\mu$, from which $F_\pi$,
in eq.~(\ref{gor}), has been calculated. In the limit of vanishing
Wilson term, this current is partially conserved and $Z_A$ reduces to
1. More generally, we find that, once the bare quark mass and the
lattice operators have been properly renormalized, the lattice
correlation functions satisfy, even in the strong coupling
approximation, the Ward identities of the continuum theory, as predicted
by recovered chiral symmetry. These identities can be then used to
derive the values of the renormalization constants of the lattice
operators. For instance, we find that the lattice conserved vector
current has indeed $Z_V=1$, and the local scalar and pseudoscalar
densities renormalize with $Z_S/Z_P=1$. The way in which this scenario
sets up is exactly the one outlined in ref.~\cite{boc} for the weak
coupling regime.

\section{The renormalized effective Lagrangian}

In the continuum limit $g_0 \rightarrow 0$ the effects of the chiral
symmetry breaking induced by the Wilson term are expected to
vanish. However, for the typical values of couplings used in current 
numerical simulations, these effects are still quite relevant, and the
Wilson terms in the lattice effective Lagrangian cannot be neglected. 

The presence of these terms complicate the task of
deriving the continuum QCD chiral Lagrangian from the lattice
calculation. First, these terms significantly increase the number of
couplings to be introduced in the lattice effective theory. Secondly,
they do not have a direct continuum limit in the QCD chiral Lagrangian,
and, because of their presence, the external sources in the lattice
effective theory do not reduce, directly, to their continuum
counterparts. Indeed, some of the effects of the Wilson terms in the
effective theory can be interpreted as a renormalization of the lattice
external sources. 

In order to overcome these difficulties, one can follow a different
procedure, which makes it possible to discard the Wilson terms in the  
lattice effective theory. The basic idea is to consider a 
``renormalized'' effective Lagrangian, which reproduces the correlation
functions of properly renormalized operators in the fundamental lattice
theory. Because these correlation functions satisfy all the Ward
identities predicted by continuum current algebra, the renormalized
effective Lagrangian does not contain the Wilson terms at all. This can
be shown explicitly in the strong coupling and large $N$ limit. In
addition, because the renormalized lattice theory exhibits the same
chiral structure of its continuum counterpart, this procedure may
significantly simplify the task of deriving the QCD chiral Lagrangian
from the lattice effective theory.

\end{document}